\documentclass[preprint2]{aastex}

\newcommand{\et}{{\it et al.}~}

\newcommand{\arcs}{\hbox{$^{\prime\prime}$}}
\newcommand{\um}{\hbox{$\mu$m}}

\shorttitle{Gorjian et al.}
\shortauthors{Infrared Nebula in NGC 5253}
\begin{document}
\title{Infrared Emission from the Radio Supernebula in NGC 5253: A 
Proto-Globular Cluster?}
\author{Varoujan Gorjian}
\affil{JPL, Caltech, MS 169-327, 4800 Oak Grove Dr. Pasadena, Ca. 91109}
\email{vg@jpl.nasa.gov}
\author{Jean L. Turner}
\affil{Department of Physics and Astronomy, UCLA, Los Angeles, Ca. 
90095-1562}
\email{turner@astro.ucla.edu}
\and
\author{Sara C. Beck}
\affil{Department of Physics and Astronomy and the Wise Observatory, Tel Aviv\\
University, Ramat Aviv, ISRAEL}
\email{sara@wise.tau.ac.il}
\begin{abstract}
Hidden from optical view in the starburst region of the dwarf galaxy NGC 5253 
lies an intense radio source with an unusual spectrum which could be interpreted variously as nebular gas ionized by a young stellar cluster or nonthermal 
emission from a radio supernova or an AGN.  We have obtained 11.7 and 18.7 \um\  images of this region at the Keck Telescope and find that it is an extremely 
strong mid-infrared emitter. The infrared to radio flux ratio rules 
out a supernova and is consistent with an HII region excited by a dense cluster 
of young 
stars. This ``super nebula" provides at least 15\% of the total
bolometric luminosity of the galaxy. Its excitation requires $10^5-10^6$ stars, 
giving it the total mass and size (1-2 pc diameter) of a globular cluster. 
However, its high obscuration, small size, and high gas density all argue that 
it is very young, no more than a few hundred thousand years old. This may be the youngest globular cluster yet observed.   
\end{abstract}
\keywords{galaxies: individual (NGC~5253)---galaxies: starburst---galaxies:
star clusters---galaxies: dwarf}
\section{Introduction}

The dwarf I0/S0 galaxy NGC 5253 is actively forming stars, with 
abundant $H\alpha$ emission, WR stars, and many optically visible star clusters 
within the central 100 pc \citep{bec96,vg96,cal97}. More than half the radio 
emission from this 
galaxy is found to come from an intense source of diameter 1.3 pc at 2 and 1.3 
cm (for a distance of 4 Mpc), in the star formation region but not identified 
with any optical counterpart \citep*{tur98}. \citet*{tur00} called this source the 
radio supernebula; it has a brightness temperature of $\rm 10^4~K$ at 2 cm and 
an optically thick spectrum that is either flat or slightly rising. 
Extra-galactic radio sources of such high $T_b$ and luminosity are usually 
non-thermal synchrotron sources with steeply falling spectra or AGN with flat 
spectra. Thermal radio sources, generally gaseous nebulae surrounding 
young stars (HII regions) have almost 
flat spectra, but are typically two to three orders of magnitude
less luminous than this radio source.  If this 
were an HII region it would require an extraordinarily high number of stars to 
excite the gas. If it were a non-thermal source it would also be 
extraordinary for its kind.  From the radio emission alone, we cannot tell if 
this is an HII region, a supernova remnant (SNR), or an Active Galactic Nucleus 
(AGN). 

We have therefore obtained high resolution images of the radio source in the 
mid-infrared, since an HII region will be a strong infrared emitter with a 
characteristic spectral energy distribution, different from a SNR or AGN. The 
observations and results are discussed in the next section, 
followed by our deductions as to the nature of the source and its properties. 

\section{Observations}

The observations were made during the re-commissioning of the Long Wavelength 
Spectrometer (LWS) \citep{jb93} on the Keck I Telescope on Mauna Kea in March 
1999.  Two 
images were obtained: one with a 1 \um\ wide filter centered at 11.7 
\um\ 
and one with a 0.5 \um\ wide filter centered at 18.7 \um.  Fluxes were 
calibrated using the standard stars $\beta$ Gem and $\alpha$ Boo and 
are estimated to have uncertainties of 10\%.  In Figure 1 we show the 11.7 \um\  
image in color superimposed on an optical photograph \citep{cal97}
from the {\it Hubble Space 
Telescope} ({\it HST}), and in Figure 2 the individual 11.7 and 18.7 \um\ 
images.  The registration uncertainty is limited by the {\it HST} image and 
is estimated to be $\sim$1\arcsec. 

The infrared source is barely resolved at a diameter 
of 0.58\arcsec\ (FWHM) at 11.7~\um\,
which is about twice the diffraction limit of the Keck.  The source is 
0.62\arcsec, only 
25\% larger than the diffraction limit, at 18.7\um. These sizes may be upper 
limits to the true source size since at the time of the observations the 
telescope settle times had not been accurately determined and inadequate settle 
times may have broadened the images. There is no significant difference between 
the appearances of the source at the two wavelengths, and allowing for the 
differences in beam size the radio and infrared images agree.  The source has 
2.2 Jy total flux at 11.7\um\, with S/N of 17 at the peak pixel and 2.9 Jy at 
18.7\um , S/N of 12 at the peak. These agree with ground-based small-aperture
10.8\um\ fluxes \citep*[$\sim$ 1.4 Jy,][]{LR79, TDW} especially when we take into account that the source is brighter at longer wavelengths. While the ISO-SWS
spectrum shows emission and absorption features within the 11.7\um\ band, our
continuum flux is in agreement with the overall
 mid-infrared continuum \citep{crow99}. IRAS fluxes 
at 12 and 25 \um\ are 2.6 and 12 Jy, respectively.
The IRAS satellite beam included the entire galaxy; the fact that the IRAS
fluxes are so close to what we measure indicates that 
this is a very bright and a very small infrared source. 

\section{The Infrared Source}

\subsection{What is It?}

There are, as we stated above, three possible sources of the  
rising-spectrum radio emission from NGC 5253: an HII
region, a SNR, or an AGN.  The discovery that the radio source is also a  
strong mid-infrared emitter narrows down the possibilities.  First, it rules out 
the presence of a bright radio supernova; supernovae are not significant 
mid-infrared sources. Most Galactic SNR are not 
detected at all in the IRAS bands \citep{are89}.   A supernova
explosion inside a dense molecular cloud will be very bright in the infrared
\citep{shu80}, but that is calculated to be a short-lived 
stage, with the emission declining significantly on a time scale of a few years 
\citep{W86}. The agreement of our measurement with the IRAS 
fluxes from 1983 and the ground-based work from 1972 and 1986 
\citep{rie72, LR79, TDW} argues against this model.

The possibility that the source is an AGN cannot be excluded so decisively.  
AGN are 
not themselves expected to be strong infrared emitters, although their spectra 
will depend on the amount and distribution of the dust nearby. Much of the 
observed range in infrared flux
 is probably due to the presence of both non-stellar central 
engines and starbursts in the same galaxy, and often in the same nuclear 
region. NGC 1068 is a nearby example of a galaxy with both 
an AGN and a powerful starburst in close quarters; at larger distances, 
such sources can be difficult if not impossible to separate \citep{he95}. The high 
spatial resolution of our observations and the closeness of NGC 5253 will 
however challenge any model which tries to combine an AGN and a starburst in the 
radio-infrared source. The conclusive test for the presence of an AGN is 
spectroscopic, not photometric, and depends on 
the detection of fast moving gas near the central engine.  That the gas 
studied optically in NGC 5253 does not have broad lines is not relevant to the 
gas in the radio source, which must be heavily obscured as it is 70 
times brighter than one would predict based on the
$H\alpha$ line \citep{cal97}.  The only 
spectroscopic information currently available on the gas inside the  radio and 
infrared source is the $10.5 \mu$m [SIV] line \citep{bec96}, which is no more 
than 60 $\rm km~s^{-1}$ wide. Those observations cannot completely 
rule out a very weak 
plateau, as from a weak AGN.  Another regime in which the influence of an 
AGN is usually strong is the 
X-ray: the X-ray spectrum and luminosity of NGC 5253 is perfectly 
consistent with a starburst in a clumpy medium \citep{mar95}. In short, while a 
small and quite unusual AGN 
could perhaps be hidden in NGC 5253, 
there is nothing in the data that needs or suggests one. 

The final possibility, which the data strongly favor, is that the source of the 
infrared and radio emission is an HII region ionized by thousands of young 
stars. We presented this model in \citet{tur00} based on radio data
alone. The high mid-infrared fluxes of the NGC 5253 
source, which exclude SNR and go far to exclude AGN, are perfectly consistent 
with the spectral energy distribution of an optically obscured HII region.  The 
ratio of $11.7\mu$m flux to the thermal radio flux is 150, a factor of 5 
higher than would be expected from pure $Ly\,\alpha$ heating \citep{ho90} and in 
agreement with results from many other star formation regions, Galactic and 
extra-Galactic. The observed ratio in starburst galaxies is usually in the 
120-250 range. 
 
The model that the NGC 5253 source is an optically obscured HII region fits {\it 
all} the data at hand: the radio spectrum, which is optically thick at 6 and 
probably 2 cm \citep{tur98, tur00}, as are dense and compact HII regions, the 
infrared flux and its ratio to the radio emission, the infrared lines which 
agree with the radio continuum \citep{bec96}, and the great excess of infrared 
and radio strengths over that predicted from the optical.  The NGC 5253 source 
could be a compact HII region such as 
W51 or K3-50 writ large, except that Galactic compact HII regions typically 
contain 1 bright star in 0.01 pc diameter; the NGC 5253 source must hold not one 
OB star but a populous cluster. The ionization requirements of the super
nebula are immense, $4\times10^{52}sec^{-1}$. The stellar population and 
density which are calculated in \citet{tur00} for a range of IMFs are 
accordingly 
high: $2\times10^5$ to $2\times10^6$ stars and
$5\times10^5$ to $10^6$ $M_\odot$ within a 1-2 pc region.  The source requires 
not 
just a cluster of stars, but a large cluster with the stellar density, the size,
and luminosity of 
a bright globular cluster.  

\subsection{On the Spectral Dominance and Spatial Concentration of the Source}

The NGC 5253 super nebula is not only remarkable in itself; it dominates the 
galaxy's total spectrum to an astonishing degree. 
80-90\% of the total galactic flux in the 12 
\um\ bands comes from this 1-2 pc region. NGC 5253 as a whole is rather blue 
in the infrared: it 
is stronger at 60$\mu$m than at 100$\mu$m and its $\nu F_{\nu}$ peaks at 
25$\mu$m; a significant fraction of the flux at 25$\mu$m may be due to a 
silicate emission 
feature \citep{crow99}. The supernebula has been imaged in only two bands, but 
also appears extremely 
blue, with a deduced dust temperature (for $\beta=1.5$) of 180 K, consistent
with the ISO mid-infrared spectrum \citep{crow99}. We do not 
know what the supernebula contributes  at 60 and 100$\mu$m, but even in the most 
extreme case that it contributes {\it nothing}, it would still 
provide $\sim$20\%, or $\rm 4\times10^8~L_\odot$,
 of the total fluxes at wavelengths below 100\um , which constitutes at least 
75\% of the total bolometric luminosity ($\rm 2.4\times10^9~L_\odot$), radio to 
X-ray, of the galaxy. We note that the population of O stars 
required for the radio emission from the supernebula \citep{tur00} will have a 
total luminosity of $0.8-1.2\times10^9L_\odot$, 
compared to $1.7\times10^9L_\odot$, the total observed
infrared luminosity. This argues that all the infrared is
distributed at the shorter wavelengths and that the super nebula gives
at least half of the 
total IR luminosity, and potentially as much as two-thirds, in an
infrared-dominated galaxy.  
So this giant HII region, less than 2 pc in diameter, may well dominate its 
host galaxy as thoroughly as does a bright AGN.  
NGC 5253 is the best example so far of 
the trend \citep{ho90} that the bulk of the infrared emission in starburst 
galaxies is usually found in a much smaller volume than the IRAS beams, and 
implied by the finding \citep{bry99, do98} that the molecular clouds which fuel star formation are concentrated in small active regions. 

\section{A Very Young Star Cluster and its Problems}

If, as we have argued, the radio-infrared source is a super nebula or giant HII 
region ionized by thousands of young  O stars, 
conditions in the source must be extreme.  If we adopt a total population of 
$10^6$ 
stars \citep{tur98,tur00} and a diameter of 1.5 pc, then the stellar separation 
is $\sim$0.02 pc. These cluster conditions, while unusual,
are actually what would be 
expected from a just-formed or still forming super star cluster or globular 
cluster.   At present the nebular density is high, $\rm 4\times10^4 
~cm^{-3}$, as deduced from the high radio optical depth \citep{tur00} and there 
is enough dust in the region to obscure the optical 
emission. As the stellar activity continues the gas and dust may be 
expected to disperse, weakening the radio and 
infrared emission and permitting the cluster to be seen in the visible, which 
will then resemble super 
star clusters which are commonly seen in starburst regions.

We argue that the source is excited by a very young super star cluster, large 
enough to become a globular cluster; just how young?  Since we do not see the 
stars we cannot find their ages directly.  The age of the nebula is  
limited from its dynamical lifetime, which must be short. 
The gas in the nebula, which is at $\sim 10^4$~K, will be overpressured 
relative to the cooler, less dense ISM and will expand at the sound speed,  
$\rm \sim 10~ km~s^{-1}$, so the 
observed supernebula should double its size in $\sim 10^5$ years. 

The problem of lifetimes, reviewed in \citet{kur00}, haunts the study of the 
Galactic ultra-compact HII regions that these young star clusters so resemble.  
Galactic ultra-compact HII regions are thought to be older than their 
dynamical ages, since the dense and clumpy
 molecular cloud environment of these sources 
slows their expansions and extends their lifetimes.  In NGC~5253 there
is no evidence for confinement by molecular gas: in fact, no CO is 
detected within 200 pc of the nebula \citep*{tbh97,mtb00}. 
Even at the low metallicity of NGC 5253, CO should be easily detectable
based on $\rm L_{IR}$.  Another way to confine the 
nebula is with the $B^2/8\pi$ pressure of the magnetic field, which will become 
comparable to the thermal pressure of the 
NGC 5253 nebulae for B fields of a few $\mu$G.  The  B field strength in
selected locations of NGC 5253 was estimated from 
minimum-energy arguments to be 50-57 $\mu$G \citep{tur98}, high for a 
galaxy but typical of a dense star-forming 
molecular cloud, and large enough that it may be 
important in 
the pressure balance of the nebula. Finally, the cluster is so massive 
and compact that it is not inconceivable that the nebula is actually
gravitationally bound, in spite of the large pressure gradient.

The question of the lifetimes of the earliest and most obscured stages of star 
formation will become increasingly important as more of these 
very young sources are observed on scales from the Galactic Ultra-compact HII 
regions to the proto-globular cluster in 
NGC~5253. Luminous obscured infrared bright sources have been seen in  
the Antennae, although larger (50 pc) than the supernebula
\citep{mir98}; and compact luminous IR sources similar to the
supernebula have been seen in NGC 253 \citep{pina92,keto93}. Both of these 
galaxies also have large numbers of young super star clusters 
\citep{wat96, whit99}. High resolution radio and CO
images of galaxies \citep{do98, kob99, tar00, bec00} 
are revealing other possible candidates, in different environments and
perhaps at different stages of evolution. There actually 
may be 
many of these young sources yet to be discovered by high resolution infrared 
observations.

\acknowledgments

We are grateful to the staff of the Keck Observatory,
and the special efforts of Randy Campbell, Fred Chaffee, and Barbara 
Schaefer in difficult circumstances.
We also thank Peter Conti, Bruce Elmegreen, Tim Heckman,
 Kelsey Johnson, Ellen Zweibel, and Dick McCray for 
helpful discussions.  JLT is supported by 
NSF grant AST 0071276 and VG thanks the National Research Council. 
 This research has made use of the NASA/IPAC Extragalactic
Database (NED) which is operated by JPL, California Institute of Technology, 
under contract with NASA.

\clearpage

\figcaption{The infrared color image is obtained from the 11.7 \um\ 
image, superimposed on the optical HST image, which is $\sim$45\arcs\
in diameter. The width of the inset is 6.5\arcsec. Orientation of the image
is NUEL. The optical HST image was provided by D. Calzetti \citep{cal97}.
\label{fig1}}

\figcaption{False color (intensity-coded) images of  11.7 \um\ (upper) and  
18.7 \um\ (lower) images from LWS on Keck I. NUEL is indicated on the 11.7\um\ image. Images are 10\arcsec\ square. 
\label{fig2}}
\end{document}